\documentclass[a4paper,12pt,fleqn]{article}

\title{Privacy sets for constrained space-filling}
\author{Eva Benkov\'{a}\hspace{.2cm}\\
 {\tt eva.benkova@jku.at}\\
    \vspace{.2cm}
    Johannes-Kepler-University Linz and Comenius University Bratislava\\
Radoslav Harman\hspace{.2cm}\\
    {\tt radoslav.harman@fmph.uniba.sk} \\
    Comenius University Bratislava\\
Werner G. M\"uller\hspace{.2cm}\\
    corresponding author: {\tt werner.mueller@jku.at}\\
    \vspace{.2cm}
    Johannes-Kepler-University Linz}

\def\P{\mathcal P}
\def\A{\mathcal A}
\def\X{\mathcal X}
\def\Y{\mathcal Y}
\def\bM{\mathbf M}

\def\bbf{\mathbf f} 
\def\bbN{\mathbb N}
\def\bbR{\mathbb R}

\usepackage{latexsym}
\usepackage{amsmath}
\usepackage{amsfonts}
\usepackage{amssymb}
\usepackage{amsthm}
\usepackage{graphicx}
\usepackage{color}
\usepackage{epsfig, epstopdf}
\usepackage{subfigure}
\usepackage{mathtools, caption}
\usepackage{enumerate}
\usepackage[lined,ruled,linesnumbered]{algorithm2e}
\usepackage{slashbox,booktabs}

\newtheorem{Definition}{Definition}

\newtheorem{Assumption}{Assumption}

\DeclareMathOperator*{\argmax}{argmax} 

\begin{document}
\renewcommand{\arraystretch}{1.2}
\maketitle

\abstract{The paper provides typology for space filling into what we call ``soft'' and ``hard''  methods along with introducing the central notion of privacy sets for dealing with the latter. A heuristic algorithm based on this notion is presented and we compare its performance on some well-known examples.}

\section{Introduction}
\label{sec:intro}
Most of the literature on space-filling designs attempts
to achieve its aim by optimizing a prescribed objective measuring a degree of space-fillingness, such as maximin, minimax, etc., sometimes combined with an estimation or prediction oriented criterion (like suggested in \cite{Morris_1995}). Let us label those as ``soft'' space-filling methods. In contrast, ``hard'' space-filling methods ensure desirable properties by enforcing constraints on the designs, such that a secondary criterion can be used for optimization (e.g. D-optimality in the case of the Bridge-designs of \cite{Jones_2014}). In this paper we intend to propose a general framework for the latter methods based on the central notion of so-called privacy sets, which allows us elegant and efficient formulations of design algorithms. Eventually we will compare our ``hard" techniques to the more established ``soft" procedures on several examples.
\bigskip

Formally, let $\X \subseteq \mathbb{R}^d$ be a non-empty design space and let $\xi$ be a finite subset of $\X$ representing an experimental design, with each point of $\xi$ being a design point of a single trial of the corresponding experiment. 
The notion of a ``set'' automatically implies that the order of design points is irrelevant, as well as that the replicated observations in the same design point are not allowed.  
Without loss of generality, the standard problem of optimal designs of experiments (cf. eg. \cite{Pazman_1986}) is to maximize some criterion  in the set of all 
 permissible designs $\Gamma$, that is, to find 
$\xi^* \in \argmax_{\xi \in \Gamma}\{\Phi(\xi)\},$
where $\Gamma \subseteq \Xi$ and $\Xi$ denotes the set of all designs. Here, the soft methods of space-filling focus on proposing and tuning the criterion $\Phi$ and then optimize on the set $\Gamma=\Xi$. Hard methods, in contrast, pose restrictions on the design, such that all designs from $\Gamma$ satisfy a required minimum level of space-fillingness.
\bigskip


The statistical quality of one design compared to another can be assessed by calculating their mutual efficiency, which is defined as $\text{eff}(\xi|\eta)=\Phi(\xi)/\Phi(\eta)$ for any $\xi,\eta \in \Xi$, $\Phi(\eta)>0$. This quantity is particularly meaningful when the criterion $\Phi$ is positively homogeneous, which means $\Phi(\alpha\xi)=\alpha\Phi(\xi)$ for any $\xi \in \Xi$ and any $\alpha \geq 0$. 

\section{Privacy sets algorithm}

Let us now define the central notion for our approach to the generation of (constrained) space-filling designs. 

\begin{Definition}
For each $x \in \X$, let $\P(x) \subseteq \X$, $x \in \P(x)$, be a given privacy set of the point $x$.
For any design $\xi \in \Xi$, let $\P(\xi)=\cup_{x \in \xi} \P(x)$ be the privacy set of the design $\xi$.
\end{Definition}

We assume that there is a given upper limit on the size of the experiment $N  \in \bbN$. A design $\xi$ will then be called permissible, if $|\xi| \leq N$  and $x \notin \P(y)$ for all $x,y \in \xi$, $x \neq y$.  A permissible design $\xi$ will be called maximal permissible if it cannot be augmented without violation of some of the constraints, i.e., if $\xi \cup \{x\}$ is not permissible for all $x \in \X \setminus \xi$.


\begin{Assumption}\label{as:maximal}
We assume that $|\xi|=N$ for any maximal permissible design.
\end{Assumption}


We would like to emphasize that the idea of privacy sets is not artificial, but can be used for example to ensure various space-filling properties. In fact, many of the widely-used designs, starting from the classical exact designs (cf. eg. \cite{Casella_2008}) restricted by at most one trial at each point, to popular Latin hypercube designs (LHD, see \cite{McKay_1979}), to Bridge designs (BD, see \cite{Jones_2014}), can be formulated in the terms of privacy sets. 
More specifically, we can set $\P(x)=\{x\}$ in the case of classical designs. For Latin hypercube designs the design space $X$ is usually a finite grid and we have $\P(x)=\{y: \exists i \in \{1,\ldots,d\}: x_i = y_i\}$, where $x,y \in \bbR^{d}$ and $x_i,y_i$ denote theirs $i$-th coordinates. Privacy sets for Bridge designs are given by equation \eqref{eq:bridge} in Section \ref{sec:PSA_BD}.
\bigskip

Note that the use of privacy sets is meaningful not only for computer, but also for physical experiments. It covers, for example, time-separation constraints, where the designs space represents time, and consecutive trials must be performed at least $\delta$ time units apart (see, e.g., the second example in Section 5 of the paper \cite{sagnol2015}). In this case, $\X=\bbR^{+}_0$ and  $\P(x)=(x-\delta,x+\delta) \cap \X$ for all $x \in \X$.
\bigskip

The set $\P(x)$ is typically some kind of a neighbourhood of $x$ (containing also $x$ itself) securing some ``privacy'' for it, although this is not strictly required by the definition itself.  
Using the privacy sets defined above we obtain the optimization problem
\begin{equation}\label{eq:argmx}
\xi^* \in \argmax_{\xi \in \Xi}\{\Phi(\xi); |\xi| \leq N$ and $x \notin \P(y)$ for all $x,y \in \xi$, $x \neq y \}.
\end{equation} 

In the following, we present a general framework of exchange-type algorithm - Privacy Sets Algorithm (PSA) - for solving optimization problem \eqref{eq:argmx}.  
In general, the specification of individual steps depends greatly on the design space $\X$ and on the constraints given by the sets $\P(x)$, $x \in \X$, as well as on the optimization criterion $\Phi$.
\bigskip

A characteristic feature of PSA is the ability to temporarily violate ``privacy'' of one or more design points. This offers a wider range  of possibilities than when performing only permissible changes and prevents the algorithm from getting stuck, for instance when the privacy constraints are very strict.
Let $\A(\xi) \subseteq \X \setminus \xi$ denote a set of ``candidate points'' that can possibly augment a maximal design $\xi$. The set $\A(\xi)$, in contrast to $\P(\xi)$, is not an attribute of the problem itself, but can be adjusted in order to ensure the optimum performance of the algorithm. Note that we do not require $\A(\xi)$ to contain solely permissible points $x \notin \P(\xi)$.          
\bigskip

PSA does not pose any restrictions on the design space $\X$. Due to implementation reasons, however, we always assume a finite design space of size $n \in \mathbb{N}$. If $n$ is relatively small (up to thousands of design points, say), the implementation of PSA is rather simple for all kinds of privacy sets. This is mainly true because of the ability to store information about availability of each individual point of the design space.
However, with $n$ increasing, it becomes computationally intensive or even unfeasible to keep $n$-dimensional vectors in the computer memory and certain specific features of a particular class of privacy sets have to be considered. 
\bigskip
 
One of the key parts of PSA is the efficient augmentation of a design that is not maximal with the remaining runs to achieve the full size $N$. This is done by employing the following forward-type procedure (Algorithm \ref{GrP}) which adds permissible design points one-by-one until a maximal design is obtained.
\bigskip

\IncMargin{1em}
\begin{algorithm}
\SetKwFunction{OptimalDesign}{OptimalDesign}
\SetKwInOut{Input}{Input}\SetKwInOut{Output}{Output}
\Input{A permissible design $\xi$, $|\xi|=N^* < N$.}
\Output{A permissible design $\xi$, $|\xi|=N$} \BlankLine
\For{$i=1:N-N^*$}{
Augment $\xi$ with the point $x$, which maximizes $\Phi(\xi \cup \{x\})$ subject to $x \notin P(\xi)$.
} 
\Return{$\xi$}

\caption{Greedy Procedure (GrP)}\label{GrP} \end{algorithm}\DecMargin{1em}

One-point permissible augmentation from Step 2 can be crucial in effective implementing of PSA algorithm. One of the straightforward solutions is to use the exhaustive enumeration of $\X \setminus \P(\xi)$ (for smaller problems) or to use a blind random search, that is, to choose the best point from a set of candidates sampled independently from $\X \setminus \P(\xi)$. This can be performed by a direct rejection method, which, however, tends to be very inefficient for some cases. Therefore, we recommend exploiting particularities of the privacy constraints, if possible (as for example in Section \ref{sec:PSA_BD}). 
\bigskip

For any maximal design $\xi$ and any point $x \in \A(\xi)$ let $\eta(\xi,x)$ be a possibly random maximal permissible design based on $\xi$, containing $x$. We can view $\{\eta(\xi,x): x \in \A(\xi)\}$ as a randomly generated neighbourhood of the design $\xi$ in the set of all maximal permissible designs, or, in other words, a set of slight ``mutations'' of the design $\xi$. 
The mutation procedure given by Algorithm \ref{MuP} calculates $\eta(\xi,x)$ for any permissible design $\xi$ and $x \in \A(\xi)$.
\bigskip

\IncMargin{1em}
\begin{algorithm}
\SetKwFunction{OptimalDesign}{OptimalDesign}
\SetKwInOut{Input}{Input}\SetKwInOut{Output}{Output}
\Input{A permissible design $\xi$, $|\xi|=N$ and a candidate point $x \in \A(\xi)$.}
\Output{A permissible design $\xi$, $|\xi|=N$} \BlankLine
Remove from $\xi$ all those points  that belong in $\P(x)$. \BlankLine
Let $\xi=\xi \cup \{x\}$. \BlankLine

\uIf{$|\xi|=N+1$}
{
Remove the design point from $\xi$ that leads to the smallest drop in the criterion value.}
\ElseIf{$|\xi|<N$}
{
Augment the design $\xi$ using GrP to the maximal design.}
\Return{$\xi$}
\caption{Mutation Procedure (MuP)}\label{MuP} \end{algorithm}\DecMargin{1em} 

The main body of the Privacy Sets Algorithm can then be written in the scheme of Algorithm (\ref{PSA}).
\bigskip

\IncMargin{1em}
\begin{algorithm}
\SetKwFunction{OptimalDesign}{OptimalDesign}
\SetKwInOut{Input}{Input}\SetKwInOut{Output}{Output}
Construct an initial design $\xi$ using GrP.\\
\Repeat{$\Phi(\xi_{old}) \geq \Phi(\xi)$}{
Set $\xi_{old} \leftarrow \xi$.\\
Construct candidate set $\A(\xi)$.\\
\For{$x \in \A(\xi)$}{
Set $\eta\leftarrow\eta(\xi,x)$ using MuP.\\
\If{$\Phi(\eta)>\Phi(\xi)$}{
Set $\xi\leftarrow\eta$.\\
break the for cycle}
}
}
\Return{$\xi$}
\caption{Privacy Sets Algorithm (PSA)}\label{PSA} \end{algorithm}\DecMargin{1em} 

\section{Privacy sets algorithm for Bridge designs}
\label{sec:PSA_BD}
In this section, we provide a particular version of PSA, which deals with so-called Bridge constraints. This term originates from  \cite{Jones_2014}, but we employ it for the constraints themselves, independent of the optimization criterion.
\bigskip

Let $\delta$ be a given positive constant. A design $\xi$ will be called Bridge design, if $\xi$ satisfies constraints given by the privacy sets defined for any $x \in \X$ by
\begin{equation}\label{eq:bridge}
\P(x)=\{y\in \X: |x_i-y_i| <\delta \textrm{ for some } i \in \{1,\ldots,d\}\}.
\end{equation}
Definition \eqref{eq:bridge} suggests that the focus of Bridge constraints is on the space-fillingness of one-dimensional projections of the design points, since no two levels of a given factor can be closer than $\delta$. Motivation for this requirement can be drawn from the non-collapsingness of the design in the case that some of the factors turn out to be irrelevant.
\bigskip

In this section, we assume $\X=[-1,1]^d$, that is, every factor takes values in a bounded interval, which can be scaled to $[-1,1]$. We assume that $\X$ is discretized into a grid of $n=L^d, L \in \mathbb{N}$, equally spaced points with each coordinate from $\left\{-1+\frac{2k}{L-1}, k=0,1,2,\ldots,L-1\right\}$. 
Without loss of generality, we will choose the minimum spacing $\delta$ from the set $\left\{ \frac{2k}{L-1}, k=1,2,\ldots \right\}$.
\bigskip 

The requirement of $N$ experimental runs implies $\delta \leq 2/(N-1)$ and $L \geq N$. Note that for the special case $L=N$, we obtain the well-known Latin Hypercube designs (LHD). If $L>N$ and $\delta= 2/(L-1)$, the resulting design can be viewed as an ``incomplete'' LHD.
\bigskip

In general, the most computationally difficult part of PSA is the one-point permissible augmentation in Algorithm \ref{GrP}. 
Using the specific nature of Bridge designs defined on $[-1,1]^d$, we implemented this step in two parts.
\bigskip

In the first part, we perform a blind random search by repeated sampling from $\X \setminus \P(\xi)$ and selecting that point which leads to the biggest increase of the criterion value. In the case of Bridge designs, it is enough to store just an $L\times d$ logical matrix representing permissible levels of factors. Points $x \in \X \setminus \P(\xi)$ can then be easily selected independently, coordinate by coordinate. In the case where additional constraints on the design space are present (see Section \ref{sec:constrX}), we used the rejection method.
\bigskip

In the second part, we tune the best point found 
by using a local search optimization procedure.
Its main idea is to sequentially improve the position of the design point added in the first part, always varying only one coordinate at a time. Since all other design points remain unchanged, we are allowed to move only in the permissible area, where no collision with another design point occurs. The process of checking feasibility of the prospective design space point can be handled easily when considering Bridge restrictions (see the reasoning above). Note that this procedure does not require storing all design points (or regressors associated with the design points) in the computer memory. Therefore PSA for Bridge designs can be applied to very large design spaces. 
\bigskip

\section{Examples: $D$-optimal Bridge designs on a cubical design space}

This section provides examples of  Bridge designs for specific choices of the optimization criterion and the design space. We compare these to the results from \cite{Jones_2014}, where the same settings were considered.

Let $\bbf(x)$ be an $m$-dimensional function at design point $x \in \xi$, $m \in \mathbb{N}$, usually representing a regression function in a linear model with uncorrelated homoscedastic errors.

In optimal experimental design, the most common objective function $\Phi$ is the $D$-criterion given by
\begin{equation}\label{eq:PhiD}
\Phi_D(\xi)=\det(\bM(\xi))^{1/m},
\end{equation}
 where 
$
\bM(\xi) = \frac{1}{N}\sum_{x \in \xi} \bbf(x)\bbf^{\top}(x)
$
is the standardized information matrix of the size $m \times m$. This definition of $D$-optimality ensures its positive homogeneity discussed in Section \ref{sec:intro}.
If $\bbf(x)$ is a regressor in a linear model, a $D$-optimal design minimizes the generalized variance of the best linear unbiased estimator of the model parameter, cf. \cite{Pazman_1986}.

This criterion leads to the optimization problem 
\begin{equation*}\label{eq:Dopt}
\xi^* \in \argmax_{\xi \in \Xi}\{\Phi_D(\xi); |\xi| \leq N$ and $|x_i-y_i| \geq\delta \textrm{ for all } i=1,\ldots,d \}.
\end{equation*} 

It is natural to require the design space to be a $d$-dimensional hypercube $[-1,1]^d$ (for example when conducting a computer experiment), hence we restrict ourselves to this assumption.

With the number of trials $N$ given, the density of the design space grid needs to be chosen. First, we know that the minimal distance $\delta$ cannot exceed $2/(N-1)$, but it is recommended to set it to a smaller value in order to maintain a certain level of ``freedom''. Now, with the value of $\delta$ determined, we can set $L=\lfloor 2k/\delta \rfloor +1$, where $k \in \mathbb{N}$ and $\lfloor . \rfloor$ represents the lower integer part. In \cite{Jones_2014} the parameter $k$ is always set to $1$, yielding the ``incomplete'' LHD defined in the previous section. In general, the choice of small $L$ can accelerate the algorithm, but also impair the quality of the resulting design, which suggests that a certain compromise should be made.
\bigskip

In the following examples, we illustrate the performance of our algorithm applied to the problems of various dimensions. We compare the results obtained by our algorithm to the results from  \cite{Jones_2014}, based on the relative efficiency of the best designs found in a fixed time interval. 
\bigskip

To make the comparison as fair as possible, we implemented both algorithms in Matlab computing environment. The algorithm of \cite{Jones_2014} was translated from JMP scripting language which was available on the supplement area of the published article.
We used 64 bit Windows 7 system running an Intel Core i3-4000M CPU processor at 2.40 GHz with 4 GB of RAM. 

\subsection{Bridge designs for 2 factors in 21 runs}
\begin{figure}[h!]
\captionsetup{width=0.9\textwidth}
\begin{center}
\subfigure[linear regressor, $\delta=0.05$]{
\resizebox*{0.4\textwidth}{!}{\label{fig:2DL05}\includegraphics{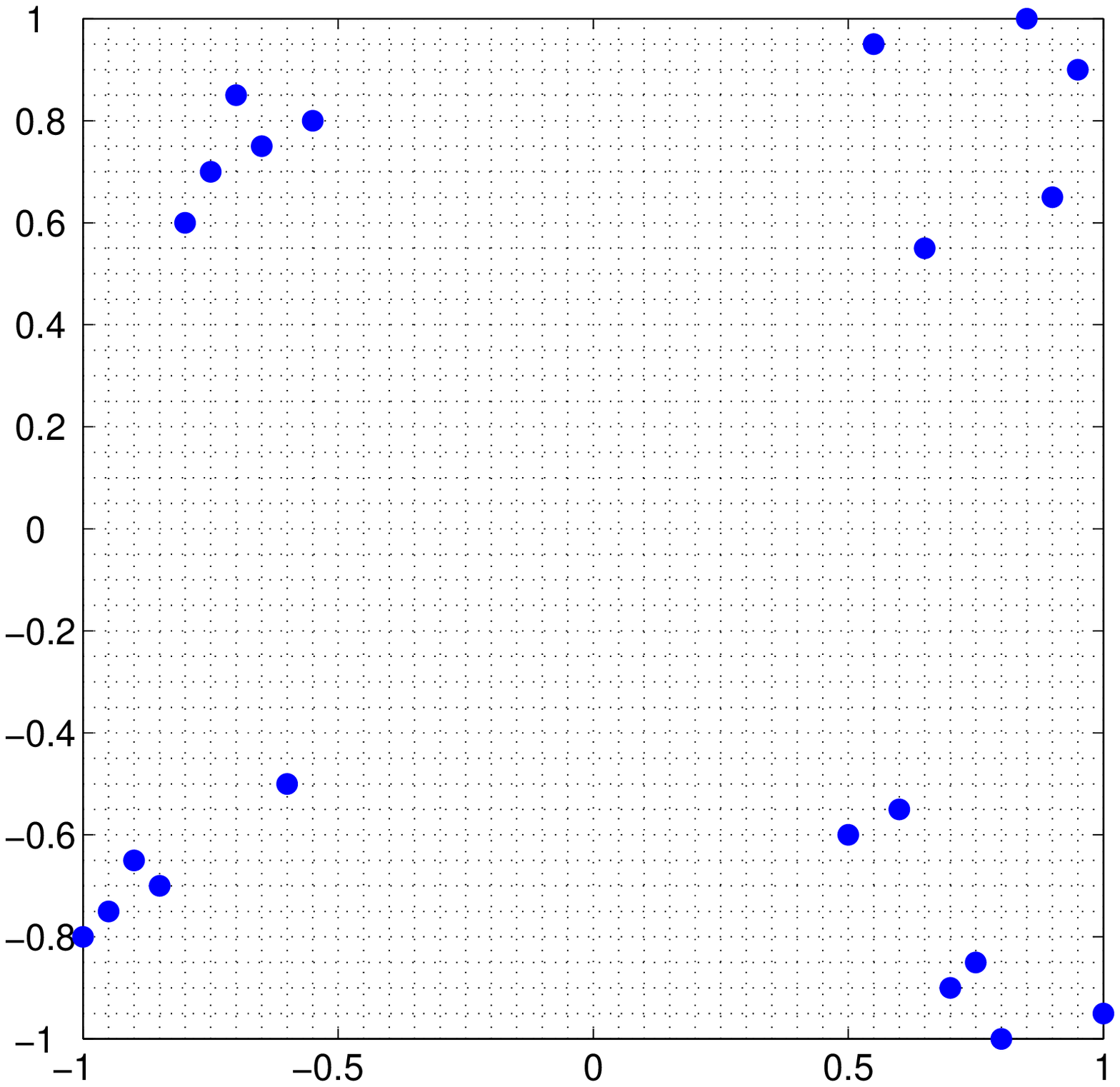}}}\hspace{25pt}
\subfigure[linear regressor, $\delta=0.025$]{
\resizebox*{0.4\textwidth}{!}{\label{fig:2DL025}\includegraphics{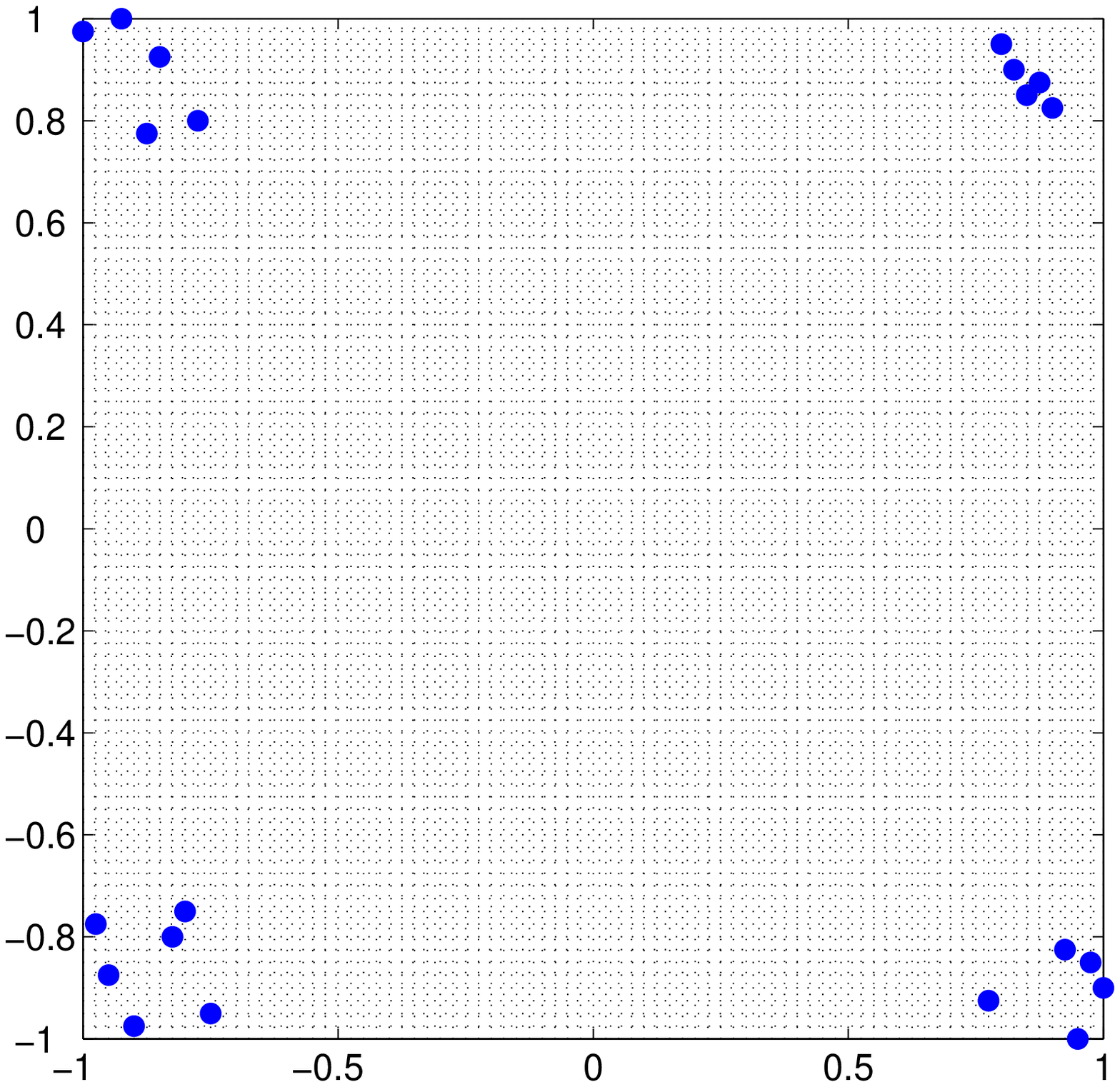}}}\\ 
\subfigure[full quadratic regressor, $\delta=0.05$]{
\resizebox*{0.4\textwidth}{!}{\label{fig:2DQ05}\includegraphics{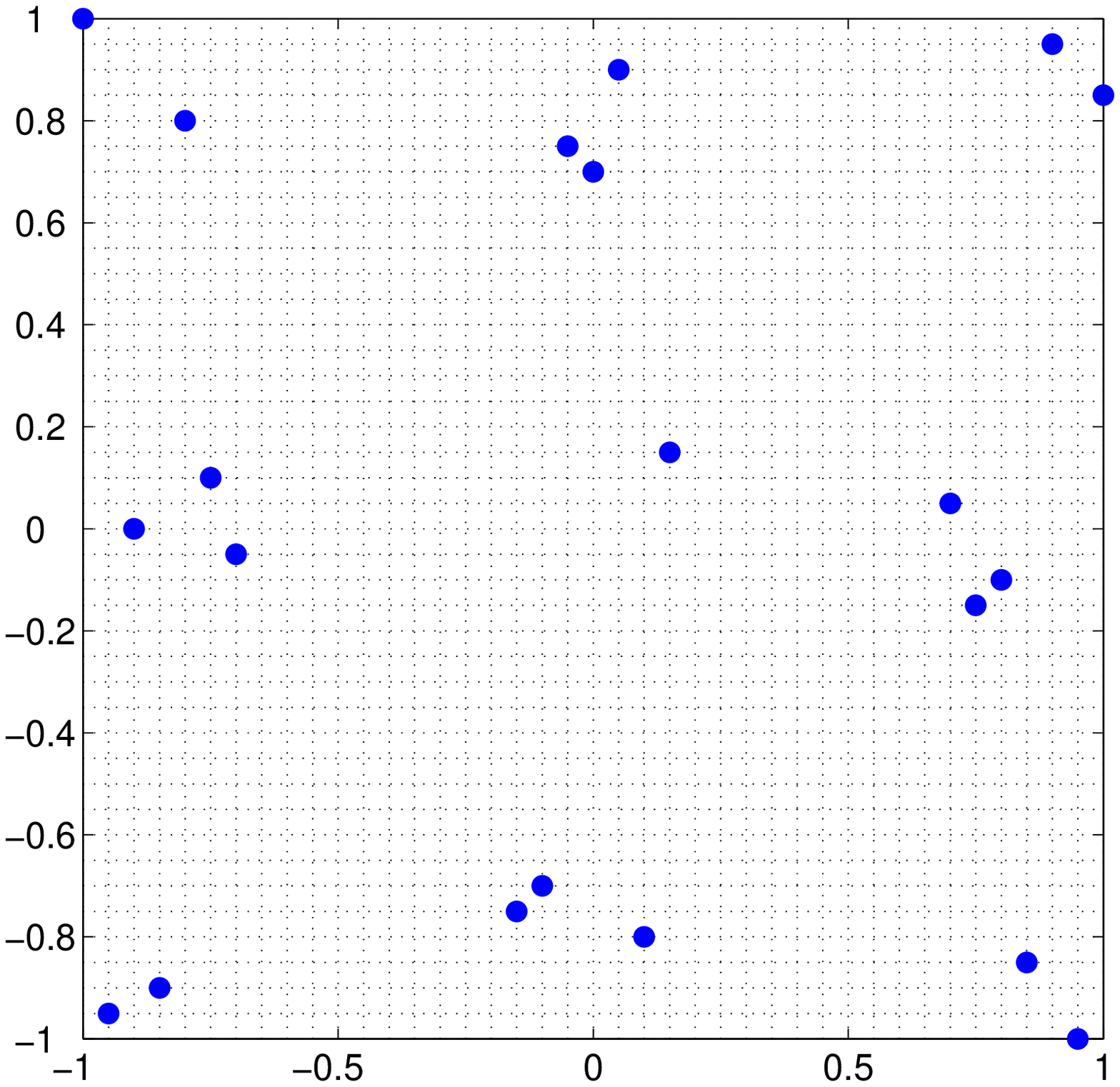}}} \hspace{25pt}
\subfigure[full quadratic regressor, $\delta=0.025$]{
\resizebox*{0.4\textwidth}{!}{\label{fig:2DQ025}\includegraphics{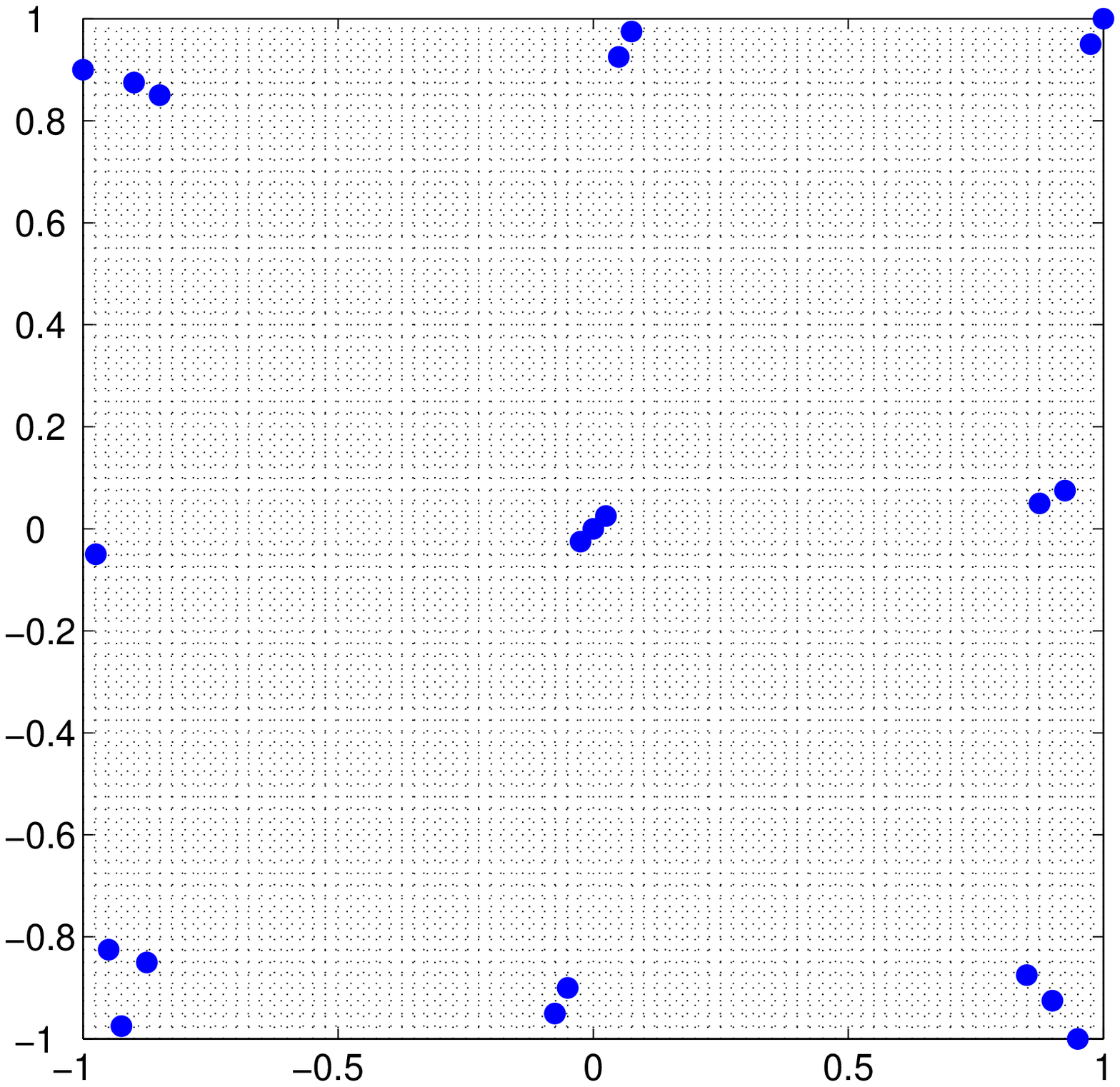}}}
\caption{Bridge designs for $N=21$ and $d=2$ found in $60$ sec. In \ref{fig:2DL05} and \ref{fig:2DL025}, the linear regressor was used and the minimal distance $\delta$ was set to $0.05$ (\ref{fig:2DL05}) and $0.025$ (\ref{fig:2DL025}). In \ref{fig:2DQ05} and \ref{fig:2DQ025} the full quadratic regressor was used and the minimal distance $\delta$ was set to $0.05$ (\ref{fig:2DQ05}) and $0.025$ (\ref{fig:2DQ025}). Efficiencies of the best designs found by the algorithm of Jones et al. relative to the designs presented in the figures \ref{fig:2DL05}, \ref{fig:2DQ05},\ref{fig:2DL025},\ref{fig:2DQ025} are $0.79$, $0.82$, $0.96$ and $0.96$, respectively.}
\label{fig:2D21N}
\end{center}
\end{figure}

As the first example, we consider the two-dimensional design space with $21$ trials to be allocated. Since this problem instance is rather small, we set the computing time to 60 seconds and run both algorithms several times, until the whole time given is spent. The designs with the highest criterion value found by PSA are displayed in Figure \ref{fig:2D21N}, with the minimum spacing constant $\delta$ set to $0.05$ and $0.025$ for both the linear regression function $\bbf(x)=(1,x_1,x_2)^{\top}$ and the full quadratic regression function $\bbf(x)=(1,x_1,x_2,x_1^2,x_2^2,x_1x_2)^{\top}$. 
\bigskip

When compared to the results of  \cite{Jones_2014}, we observe better results of our algorithm in all four examined situations. The mutual efficiencies of the designs of  \cite{Jones_2014} relative to our designs presented in the figures \ref{fig:2DL05}, \ref{fig:2DL025}, \ref{fig:2DQ05}, \ref{fig:2DQ025}  are equal to $0.79$, $0.82$, $0.96$, and $0.96$, in the respective order.  

This suggests that our algorithm provides better results especially when the spacing constraint $\delta$ is rather large, which is not surprising. 
First, the finer design space grid automatically makes the relative difference in the output designs less significant. Second, the greater the value of $\delta$ is in comparison to its upper bound $2/(N-1)$, the less space there is for ``manoeuvering'' for the algorithm. For example, the marginal case $\delta = 2/(N-1)$ leads to the standard LHD, where no observation can be added without violating some of the privacy sets constraints. These restrictions would fully disable the coordinate-exchanges in the algorithm of \cite{Jones_2014}, but could still be handled by PSA.

\subsection{A numerical study on Bridge designs}
The comparison of the algorithm of \cite{Jones_2014} and PSA can be extended into more-dimensional cases as well. We performed a small comparative numerical study on a few examples of quadratic regression of various dimensions, similar to the examples provided in Sections 3 and 4 of \cite{Jones_2014}. 
For every example, we ran the algorithms for a restricted time, observing the time dependence of the criterion value of the actually best design found by the algorithm. We repeated this procedure several times in order to provide responses from multiple random starts.

\begin{figure}[h!]
\captionsetup{width=0.9\textwidth}
\begin{center}
\subfigure[$d=2$, $N=21$, $T=10$, $t=0.05$]{
\resizebox*{0.4\textwidth}{!}{\includegraphics{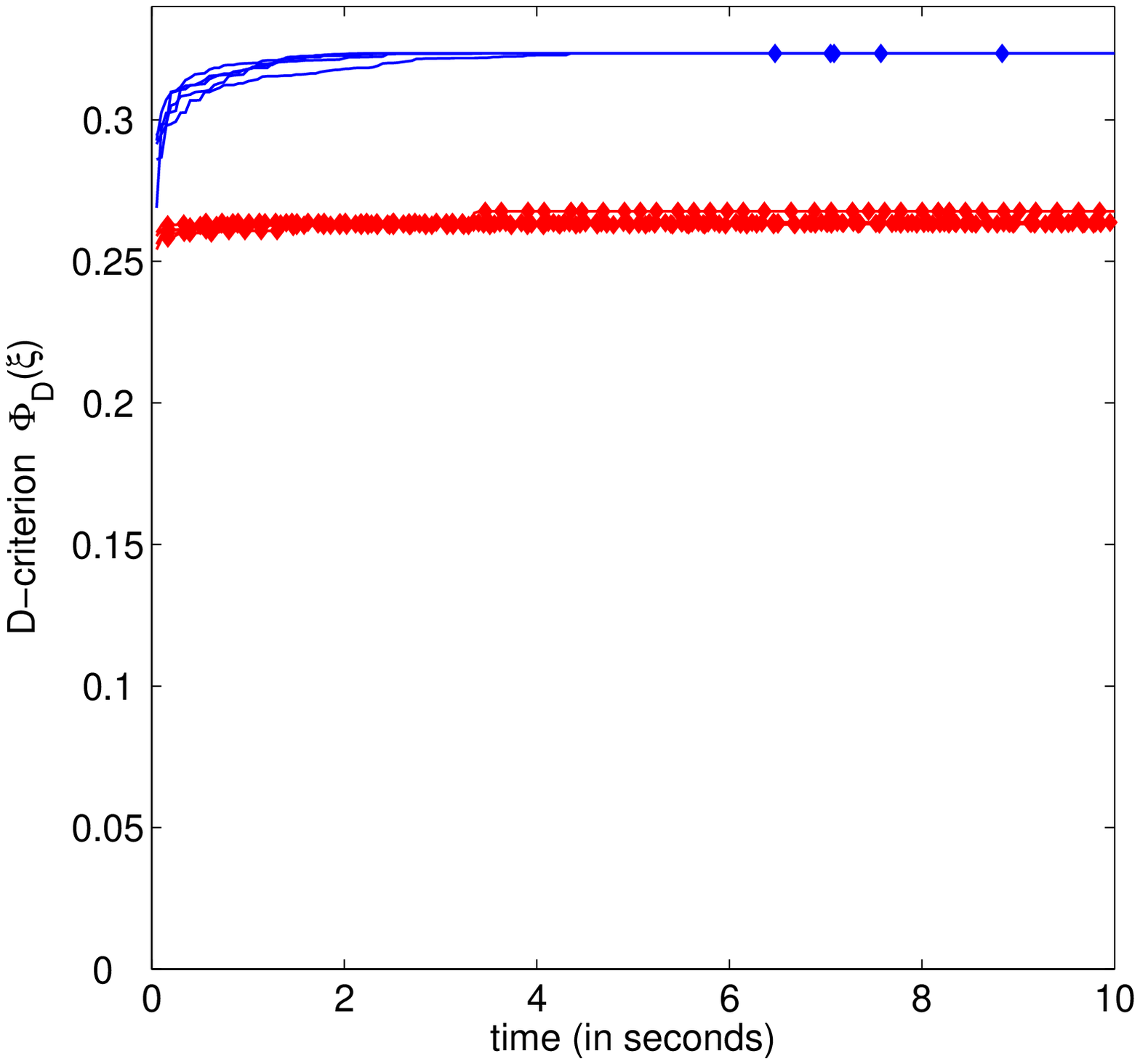}}}\hspace{25pt}
\subfigure[$d=4$, $N=41$, $T=800$, $t=2$]{
\resizebox*{0.4\textwidth}{!}{\includegraphics{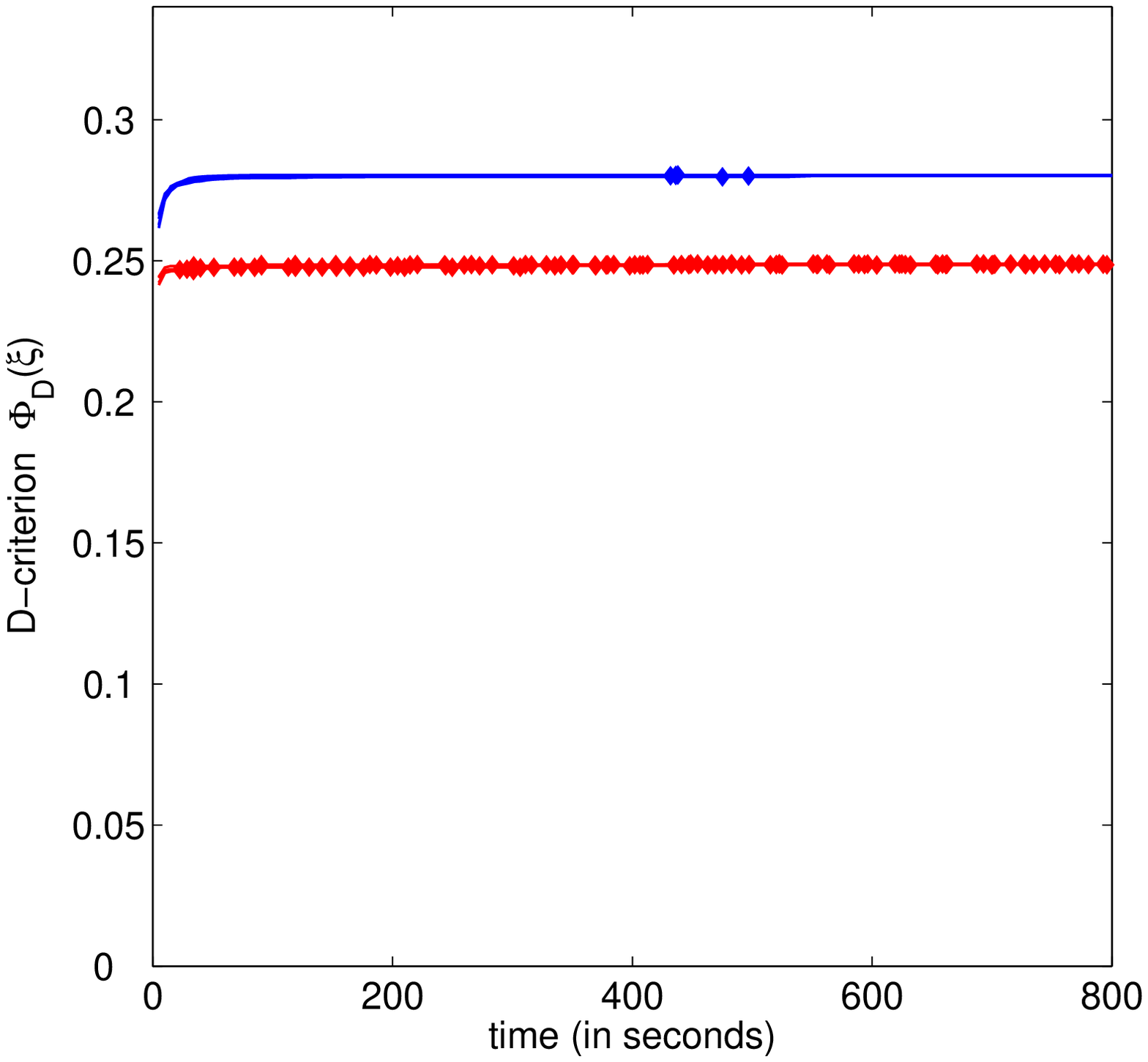}}}\\ 
\subfigure[$d=6$, $N=61$, $T=3600$, $t=5$]{
\resizebox*{0.4\textwidth}{!}{\includegraphics{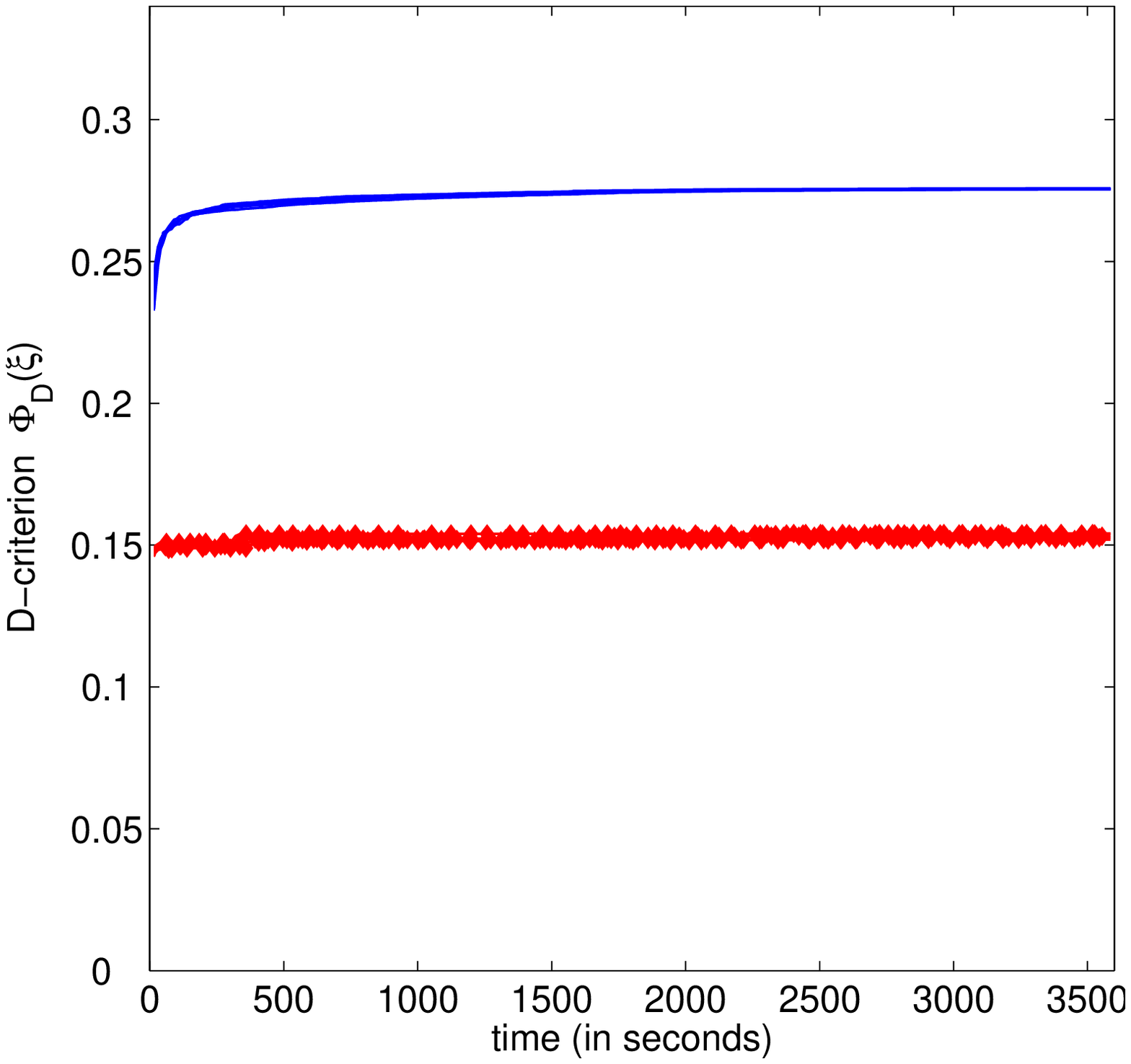}}} \hspace{25pt}
\subfigure[$d=8$, $N=81$, $T=7200$, $t=10$]{
\resizebox*{0.4\textwidth}{!}{\includegraphics{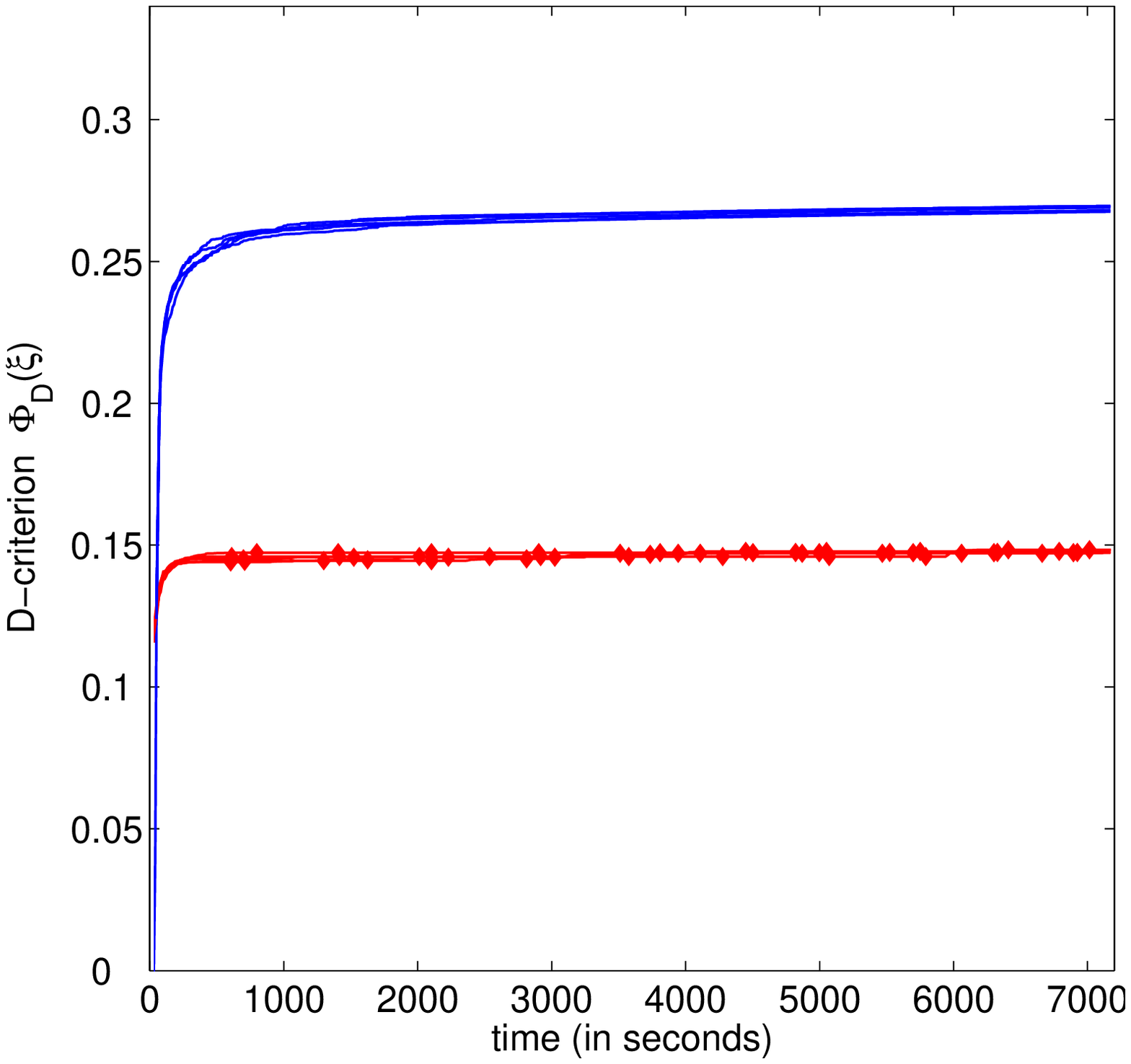}}}
\caption{Comparison of the performance of the algorithm of \cite{Jones_2014} (red lines) and PSA (blue lines). Y-axis shows the best $D$-optimality values found by the time displayed on the x-axis (in seconds). The red and the blue diamonds denote restarts of the corresponding algorithms. For every example both algorithms were ran $5$ times for the time period $T$.}
\label{fig:BD_numstudy}
\end{center}
\end{figure}

In Figure \ref{fig:BD_numstudy}, we present results of the two competing algorithms for dimensions $d=2,4,6,8$ with the numbers of trials $N=21,41,61,81$, respectively. The minimum spacing $\delta$ was in all four cases set to the value $\delta=1/(N-1)$, which corresponds to the value recommended in  \cite{Jones_2014}. 
Every $t$ seconds, we plotted the criterion values of the best designs found by the algorithm of \cite{Jones_2014} (represented by the red lines) and PSA (represented by the blue lines) versus the time displayed on the x-axis. Both algorithms were restarted $5$ times, yielding $5$ red and $5$ blue lines for each problem instance. 

The total computational time $T$, as well as the time period $t$, were chosen such that they increase with the increasing size of the problem. If an algorithm terminated during the given time $T$, it was  automatically restarted  and the resulting value found by its  run was stored in the memory. These restarts are denoted by the red and the blue diamonds. We considered the $D$-optimality criterion given in \eqref{eq:PhiD} and plotted its values on the y-axis.

Note that PSA was able to significantly outperform the algorithm of  \cite{Jones_2014} in all four examined cases. We also note that PSA yielded an efficient design in a relatively short time, which suggests that it can be reasonable to stop PSA before reaching an actual local optimum and reduce the execution time.

Although the $D$-criterion is of central interest in this section, one could be possibly interested in space-filling properties of the resulting designs, as well. There exist plenty of space-filling criteria to choose from when comparing various designs (e.g., recall that Bridge constraints themselves force a certain one-dimensional space-fillingness). We choose one of them, provided in \cite{Jones_2014}, which is based on the distances of the selected points to the nearest design point.

\begin{figure}[h!]
\captionsetup{width=0.9\textwidth}
\begin{center}\includegraphics[width=0.6\textwidth]{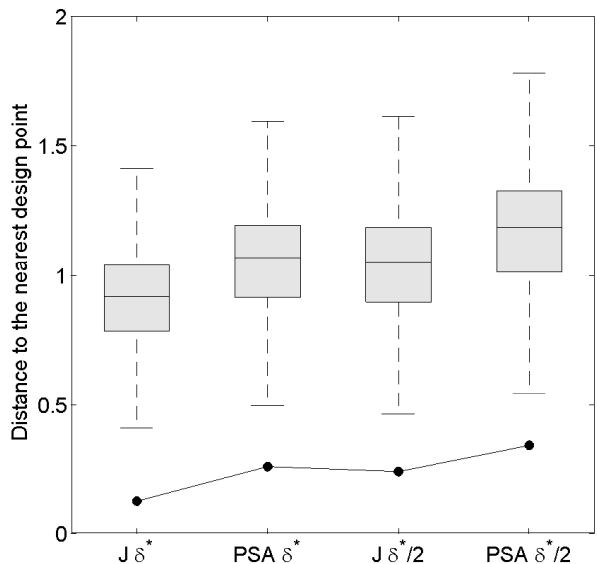}
\caption{Box plots of distances of $10000$ randomly selected points and $64$ vertices to the nearest design points for $d=6$ and $N=32$. The presented designs are divided according to the minimal distance ($\delta^*$ or $\delta^*/2$) and the algorithm used (\cite{Jones_2014} or PSA). The black dots in the bottom part represent the $D$-criterion values. Although the results of \cite{Jones_2014} are worse in terms of $D$-optimality, they are slightly better in terms of space-fillingness.}
\label{fig:BD_boxplots}
\end{center}
\end{figure}

In Figure \ref{fig:BD_boxplots} we study $4$ different $32$-point designs in $6$ factors. For every design, we display a box plot of distances to the nearest design point of $64$ vertices and $10 000$ points randomly sampled  from $[-1,1]^d$. The designs compared are gained either by the algorithm of \cite{Jones_2014}(``J'') or by Privacy sets algorithm (``PSA''), with the minimum spacing $\delta$ set to either $\delta^*=1/(N-1)=1/31$ or $\delta^*/2=1/62$. 

The lower part of Figure \ref{fig:BD_boxplots} shows $D$-optimality values of the displayed designs, denoted by the black dots. Roughly speaking, we can say that the behaviour of these values approximately follows the behaviour of the corresponding box plots. In other words, this means that the design which is the best according to the distances to the nearest design point (``J $\delta^*$''), has the worst value of $D$-criterion, and vice versa. This is not surprising, since $D$-optimal designs have the tendency to concentrate in a few points of the design space and do not possess any particular space-filling properties. 


\section{Space-filling designs on a constrained design space}

\label{sec:constrX}
Note that we can practically think of any space-filling design as a particular instance of a `Bridge design' and that this does not necessarily include $D$-optimality and a cubical design region. In fact, for any such design, it is enough to satisfy \eqref{eq:bridge}, that is, restrictions ensuring non-collapsing properties of the design. As an example, we present in this section space-filling on a constrained design space.
\bigskip

For that purpose, we choose one of the space-filling criteria to be optimized, which means that we combine together `soft' and `hard' methods described in Section \ref{sec:intro}. We consider the  average reciprocal distance (ARD) criterion, modified such that the optimal design has good projection properties onto a given set
of subspaces of the design space (as introduced in \cite{Draguljic_2012}).
Let $J\subset \{1,2,\ldots,d\}$ be a nonempty index set of dimensions of subspaces we would like to consider and let $\X_j$ denote the set of all $\binom{d}{j}$ standard coordinate subspaces of dimension $d$ 
for every $j \in J$.
\bigskip

The idea of the ARD criterion is that the average reciprocal pairwise distances of design points should be minimized. Hence, in order to remain consistent with the `maximization policy' adopted in this paper (see equation \eqref{eq:argmx}), we use the following formulation of ARD:

\begin{equation}\label{eq:ARD}
\Phi_{ARD}(\xi)=\left(\frac{1}{{N \choose 2} \sum_{j \in J}{d \choose j}} \sum_{j \in J} \sum_{\Y \in \X_j} \sum_{\substack{x,y \in \xi \\ x\neq y}} \left(\frac{j^{1/z}}{\rho_z(x^*_{\Y},y^*_{\Y})}\right)^{\lambda}\right)^{-1/\lambda},
\end{equation}
where $z \geq 1$ and $\lambda \geq 1$ are given constants, $x^*_{\Y}$ is the projection of $x \in \X$ onto subspace $\Y$, and $\rho_z$ is for any couple $x=(x_1,\ldots,x_d)^{\top}$, $y=(y_1,\ldots,y_d)^{\top}$ defined by
\begin{equation*}
\rho_z(x,y)=\left(\sum_{i=1}^d |x_i-y_i|^z\right)^{1/z}.
\end{equation*}

Each design point has to be selected from a design space, which in this case will be some linearly constrained region.  However, the PSA algorithm for Bridge designs on cubical regions, described in Section \ref{sec:PSA_BD}, can be rather straightforwardly adapted for the constrained design regions. 
\bigskip

Without loss of generality, assume that the design space $\X$ is a subset of $[-1,1]^d$ with some additional linear constraints $A x \leq b$ to be satisfied for every $x \in \X$, where $A$ is an $k \times d$ matrix , $k \in \bbN$, and $b \in \bbR^d$. For a given number of trials $N$, let us have the design space discretized by first making a grid of $L^d$ points on $[-1,1]^d$ (see Section \ref{sec:PSA_BD}), and then accepting only those satisfying $A x \leq b$. The parameter $L$, determining the density of the grid, has to be chosen large enough, such that not only a permissible design $\xi \in \X$ exists, but also that Assumption \ref{as:maximal} is satisfied.
\bigskip

The only question is how to implement the one-point permissible augmentation from Algorithm \ref{GrP}. We utilize blind random search on the set $\X \setminus \P(\xi)$ and select the best point found. In the case of a Bridge design on $[-1,1]^d$, we have a simple tool to sample from  $\X \setminus \P(\xi)$, by keeping track of permissible and non-permissible  levels of individual factors, see Section \ref{sec:PSA_BD}. If additional linear restrictions are present, $x \in \X \setminus \P(\xi)$ can be generated in the same way and then simply accepted if the condition $A x \leq b$ holds and rejected otherwise. Clearly, effectiveness of this rejection method depends on the design region defined by the constraints, as well as on the dimension $d$. In every step, we have to check the restriction on the design space, but we do not have to check  the collision with other design points in terms of privacy sets (due to the convenient Bridge constraints), which makes the rejection method more efficient than in the case of general privacy sets.
\bigskip

\begin{figure}[h!]
\captionsetup{width=0.9\textwidth}
\begin{center}
\subfigure[$J=\{1\}$]{
\resizebox*{0.3\textwidth}{!}{\includegraphics{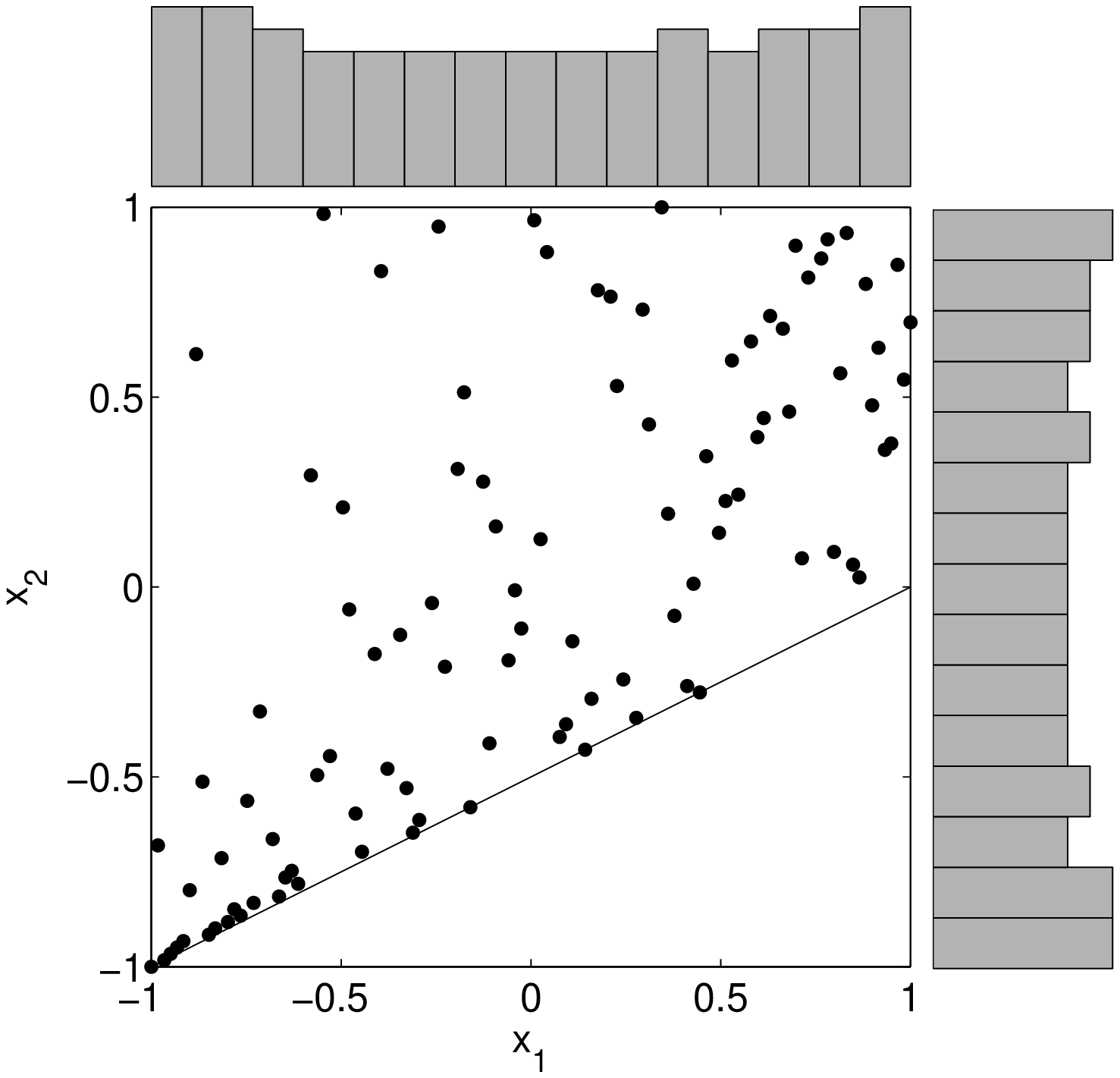}}\label{fig:BD_ARD2dJ1}}
\subfigure[$J=\{2\}$]{
\resizebox*{0.3\textwidth}{!}{\includegraphics{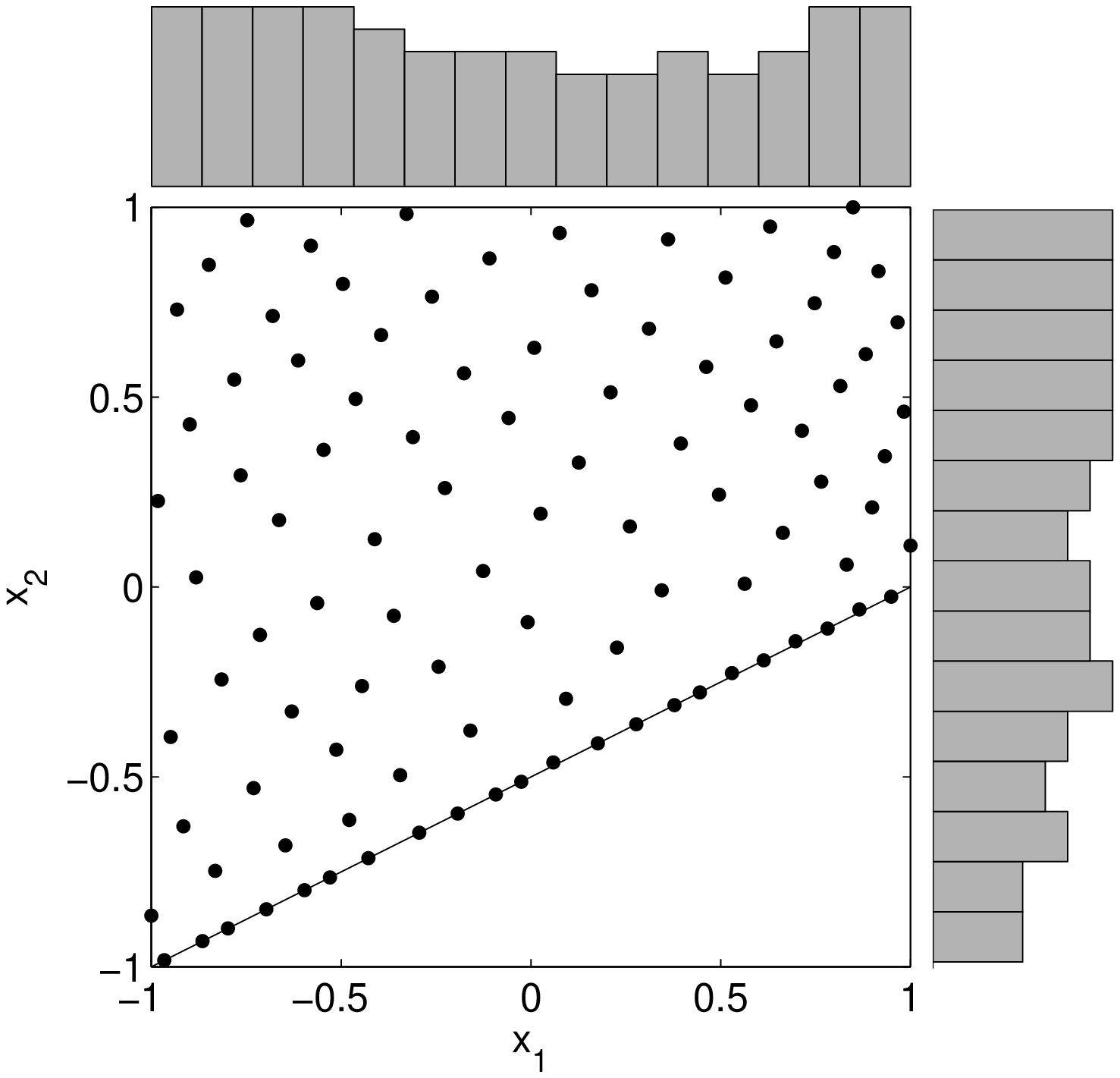}}\label{fig:BD_ARD2dJ2}}
\subfigure[$J=\{1,2\}$]{
\resizebox*{0.3\textwidth}{!}{\includegraphics{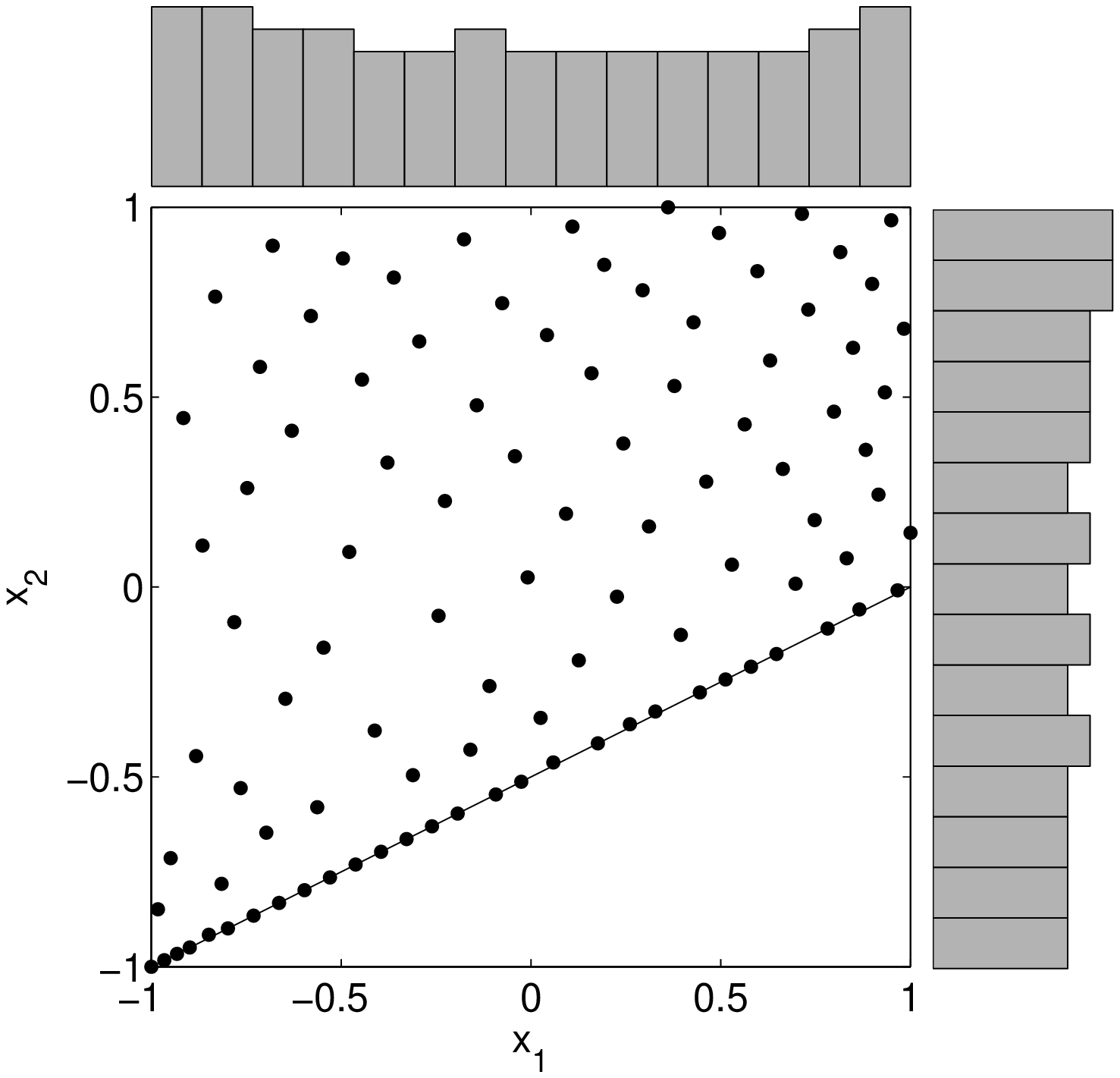}}\label{fig:BD_ARD2dJ12}} \hspace{25pt}
\caption{Designs on the two-dimensional constrained design region, obtained by PSA optimizing the ARD criterion for $z=1$, $\lambda=1$ and different values of $J\subseteq \{1,2\}$. PSA allocated $100$ trials for the minimum spacing constant $\delta=\frac{2}{120-1}$. }
\label{fig:BD_ARD2d}
\end{center}
\end{figure}

Figure \ref{fig:BD_ARD2d} displays three resulting designs obtained by PSA for different variants of the ARD criterion. The design space is in all three situations the square $[-1,1]^2$ additionally restricted by $\frac{1}{2} x_1 - x_2 \leq \frac{1}{2}$. We considered designs with $N=100$ runs and the minimum spacing parameter $\delta=\frac{2}{120-1}$. We note that in this case, it is not possible to set $\delta=\frac{2}{N-1}$ (``LHD-setting''), since Assumption \ref{as:maximal} would not be fulfilled and the PSA algorithm could not be executed.
\bigskip

The parameters of the ARD criterion \eqref{eq:ARD} were set to $z=1$ and $\lambda=1$. The nonempty index set $J \subseteq \{1,2\}$ varies in figures \ref{fig:BD_ARD2dJ1}, \ref{fig:BD_ARD2dJ2}, \ref{fig:BD_ARD2dJ12}  through all three possibilities, leading to different output designs. Space-filling properties of these designs for various settings of ARD criterion are compared in Table \ref{tab:BD_ARD2d}.
\bigskip

\begin{table}[htbp]
\captionsetup{width=0.9\textwidth}
\centering
\begin{tabular}{llll}
\toprule
\textbf{Design} &\makebox[4em]{Fig. \ref{fig:BD_ARD2dJ1}}&\makebox[4em]{Fig. \ref{fig:BD_ARD2dJ2}}&\makebox[4em]{Fig. \ref{fig:BD_ARD2dJ12}}
\\
\midrule
\textbf{Criterion} &&&\\
ARD $J=\{1\}$ & \makebox[4em]{$4.2497$} & \makebox[4em]{$4.4106$} & \makebox[4em]{$4.2718$}\\
ARD $J=\{2\}$ & \makebox[4em]{$2.6351$} & \makebox[4em]{$2.1876$} & \makebox[4em]{$2.2473$}\\
ARD $J=\{1,2\}$ & \makebox[4em]{$3.7115$} & \makebox[4em]{$3.6696$} & \makebox[4em]{$3.5970$}\\
\bottomrule
\end{tabular}
\caption{Space-filling properties of designs displayed in Figure \ref{fig:BD_ARD2d} measured by ARD criterion for different values of $J\subseteq \{1,2\}$.}
\label{tab:BD_ARD2d}
\end{table}

Histograms in Figure \ref{fig:BD_ARD2d}, as well as the first row of Table \ref{tab:BD_ARD2d}, measure space-fillingness of one-dimensional projections of the designs. Ideally, we would like to have these projections as uniformly distributed as possible (corresponding to Latin Hypercube constraints), which would ensure non-collapsing properties in one-dimension. One could quite easily think of a heuristics forcing perfectly uniform one-dimensional projections of a design. However, we think it can be more beneficial to ``relax'' constraints from LHD to BD for some smaller value of $\delta$, if this allows us to improve another optimization criterion, e.g. the criterion assessing space-filling properties of more-dimensional projections (ARD for $J=\{2\}$ or $J=\{1,2\}$). In this sense, we can compare designs of Figure \ref{fig:BD_ARD2d} to the design from the paper \cite{Petelet2010}, given in Figure 4 (a) of Section 2.3, which differs only in the scaling of the design space. Although the design of \cite{Petelet2010} strictly satisfies Latin Hypercube constraints (see the histograms of one-dimensional projections), it completely ignores space-fillingness in other dimensions.
\bigskip

\begin{figure}[h!]
\captionsetup{width=0.8\textwidth}
\begin{center}\includegraphics[width=0.9\textwidth]{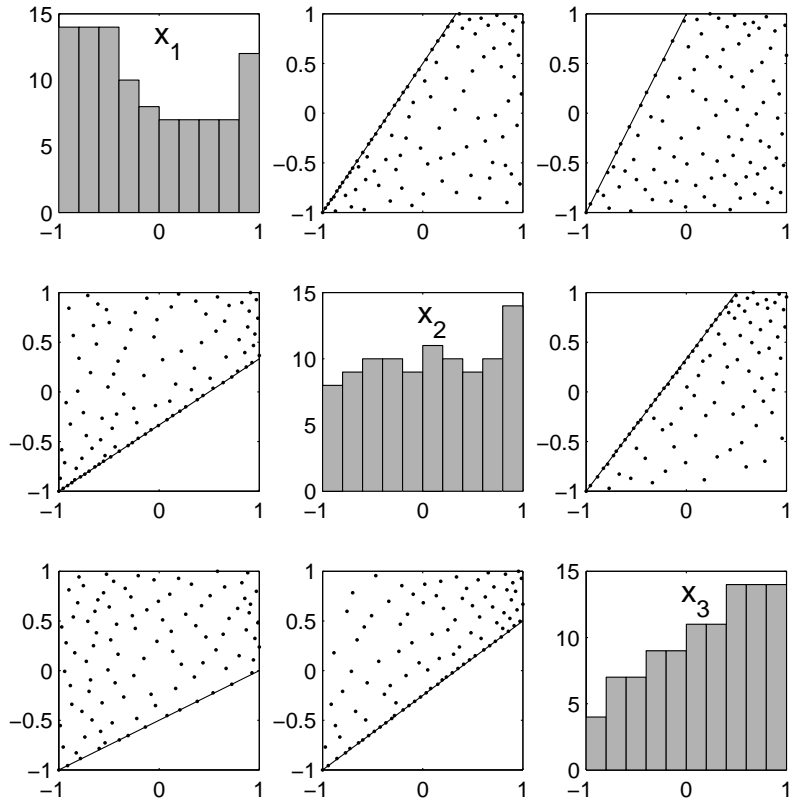}
\caption{Design on the three-dimensional constrained design region, obtained by PSA optimizing the ARD criterion for $z=1$, $\lambda=1$ and $J= \{2,3\}$. PSA allocated $100$ trials for the minimum spacing constant $\delta=\frac{2}{140-1}$. }
\label{fig:BD_ARD3dJ23}
\end{center}
\end{figure}

Figure \ref{fig:BD_ARD3dJ23} presents an example of a design resulting from PSA for ARD criterion in three factors. The three dimensional design space cube $[-1,1]^3$ is additionally constrained by $\frac{2}{3} x_1 - x_2 \leq \frac{1}{3}$ and $\frac{3}{4} x_2 - x_3 \leq \frac{1}{4}$. The design consists of $N=100$ trials with their one-dimensional projections not closer together than $\delta=\frac{2}{140-1}$. The ARD criterion was employed for $z=1$, $\lambda=1$ and $J=\{2,3\}$, which means that we combined the `hard' method forcing one-dimensional space-fillingness and the `soft' method ensuring space-filling properties for dimensions $2$ and $3$.

\section{Conclusions}

As we believe to have shown in this paper the notion of privacy sets is central to the understanding and interpretation of ``hard'' space-filling. The algorithms based on this notion are extremely flexible and can be used in a great variety of situations. As we have also shown we are not restricted to the ``hard'' space-filling paradigm, but we can encompass ``soft" methods and combinations as well.

This is perhaps emphasized best by relating to the recently proposed MaxPro criterion of \cite{Joseph+al_2015}, which was  designed for balancing out performance in all dimensions and is given by
\begin{equation}\label{eq:ARD}
\Phi_{MaxPro}(\xi)=  \sum_{\substack{x,y \in \xi \\ x\neq y}} \left(\frac{1}{\Pi_{i=1}^d \|x^*_i - y^*_i \|^z}\right),
\end{equation}
where usually $z=2$. In the discussion of \cite{Joseph_2015} we have demonstrated that our techniques can be easily adapted for these purposes. These and other useful extension will be a matter of our further research.

\section*{Acknowledgements}
The research of the first author has been financially supported by the ANR/FWF grant I-833-N18, the research of the second author was supported by the VEGA 1/0163/13 grant of the Slovak Scientific Grant Agency.


\end{document}